\documentclass[a4paper,11pt]{article}
\usepackage{pos}

\usepackage[mathlines]{lineno} 

\usepackage{subcaption}

\title{Searches for unusual signatures from dark sectors with the ATLAS Experiment}
\ShortTitle{Searches for unusual signatures from dark sectors with ATLAS}

\author*[a]{Danielle Wilson-Edwards} 

\onbehalf{on behalf of the ATLAS Collaboration}

\affiliation[a]{School of Physics and Astronomy, University of Manchester\\ Manchester, United Kingdom}

\emailAdd{danielle.joan.wilson@cern.ch}

\abstract{Various theories beyond the Standard Model predict unusual signatures, including new, long-lived particles decaying at a significant distance from the collision point. These unique signatures are difficult to reconstruct and face unusual and challenging backgrounds. This contribution will focus on recent results from searches for unusual signatures motivated by dark sector models, utilising pp collision data collected by the ATLAS experiment at the Large Hadron Collider. }

\FullConference{
    Presented at the 32nd International Symposium on Lepton Photon Interactions at High Energies, Madison, Wisconsin, USA, August 25--29, 2025
    }

\begin{document}

\maketitle



\section{Introduction}

Evidence of Dark Matter (DM) is abundant (for example, see Refs. \cite{Clowe_2006, 1980Rubin_DM, Persic:1995ru, 2018_des_collab_results, planck_collab_2021}). Its interactions with the Standard Model (SM) particles, however, remain unestablished. Many extensions to the SM attempt to describe DM’s particle nature and interactions. One such extension is the dark sector (DS) \cite{Strassler_2007, Strassler_2008, Han_2008}, comprising new particles that interact with the SM only via a mediator that couples to both sectors. These mediators include dark photons, sterile neutrinos, axions, or new scalar or vector bosons. The ATLAS \cite{ATLAS:2008xda} collaboration has conducted searches for different DS scenarios at the Large Hadron Collider (LHC) \cite{LHC_Evans:2008zzb}, which can manifest as a wide variety of experimental signatures. For instance, final states might be invisible or partially invisible, contain displaced decays, or feature jets with atypical topologies. Since the ATLAS detector was optimised to detect and identify SM particles which are produced in the prompt decay of SM or BSM particles, non-standard events could be rejected by standard triggers or default reconstructions. Consequently, these searches push the limits of detector performance by employing novel techniques such as dedicated triggers, special reconstructions, and data-driven background estimation. These proceedings present two unique signatures that probe different DS models.


\section{Search for emerging jets}

The ATLAS collaboration searched for emerging jets \cite{EJ_ATLAS:2025bsz} using 51.8 fb $^{-1}$ of proton–proton collision data at $\sqrt{s} = 13.6$ TeV, collected during the LHC Run 3 (2022-2023). The search targets a DS with QCD-like dynamics. In such models, ``dark quarks'' are charged under a new confining SU$(N_\mathrm{D})$ gauge group, where $N_\mathrm{D}$ is the number of dark colours, and couple to SM quarks via a new mediator particle. Dark quarks hadronise into dark mesons (namely, the dark pion $\pi _\mathrm{D}$, and dark
vector meson $\rho _\mathrm{D}$), whose lifetimes determine the decay modes: they may promptly decay back to SM particles (dark jets), remain stable and invisible in the detector (potentially providing a DM candidate), or be metastable and travel some distance before decaying, producing displaced vertices within the detector. The latter case is the focus of this analysis, where long-lived dark mesons produce emerging jets \cite{EJ_Schwaller_2015} containing multiple secondary vertices.

This search targets emerging jets from two different mediator scenarios: an s-channel process via a vector boson $Z '$ that couples to SM and dark quarks separately, and a t-channel process via a scalar $\Phi$ that couples simultaneously to one SM quark and one dark quark. Jets are classified using two complementary analysis strategies: namely a cut-based strategy to increase interpretability and a machine-learning (ML) based strategy to optimise signal sensitivity.  Events are required to have two large-radius jets, and classification is performed in two dijet invariant mass ($m_{\mathrm{jj}}$) regions, each with a trigger: $m_{\mathrm{jj}}<1$ TeV events use a dedicated emerging-jet trigger that exploits the jet's prompt track $p_\mathrm{T}$ fraction (PTF), while $m_{\mathrm{jj}}>1$ TeV events use a high-$p_{\mathrm{T}}$ single-jet trigger. Dark pion proper decay lengths between $1$ and $1000$ mm are considered, ensuring that most decays occur within the inner detector.
The dominant background from QCD multijet events is estimated using two independent data-driven methods, corresponding to each tagging strategy.

The cut-based analysis identifies emerging jets using their displaced and multi-pronged structure. In both low- and high-$m_{\mathrm{jj}}$ regions, the leading and subleading jets must each contain at least one displaced vertex (DV). In the high-
$m_{\mathrm{jj}}$ region, PTF and the 2-point energy correlation function (ECF2)\cite{ECF_Larkoski_2013}, which measures the jet’s angular energy dispersion, are used to select multi-pronged emerging jets. In the low-$m_{\mathrm{jj}}$ region, where PTF is already constrained by the emerging jet trigger, selection is based on number of sub-jets within the large-radius jet. The background yield in the signal region is estimated using the ABCD method.


The ML based strategy uses a transformer-based jet tagger, based on the ATLAS GN2 flavour tagging algorithm \cite{atlascollaboration2025transformingjetflavourtagging}, which takes features from the jet and up to 200 associated tracks. The model simultaneously optimises three tasks: classifying the jet as emerging or not, identifying track origins, and predicting whether track pairs originate from the same vertex. The jet classification score is used to tag emerging jet candidates, with a high threshold defining the signal region. Events with two or more tagged jets define the signal region. Correlations between jet features make the standard ABCD method unsuitable for background estimation. Instead, a data-driven approach based on the mistag rate - or, the probability of a background jet being tagged as signal - is used. The mistag rate is measured in the control regions and parameterised in two dimensional bins of the dominant correlated jet properties, $p_\mathrm{T}$ and PTF. These mistag rates are then used to predict the background yield in the signal region. 

\begin{figure}[h!]
    \centering
    \begin{subfigure}[b]{0.35\textwidth}
        \centering
        \includegraphics[width=\textwidth]{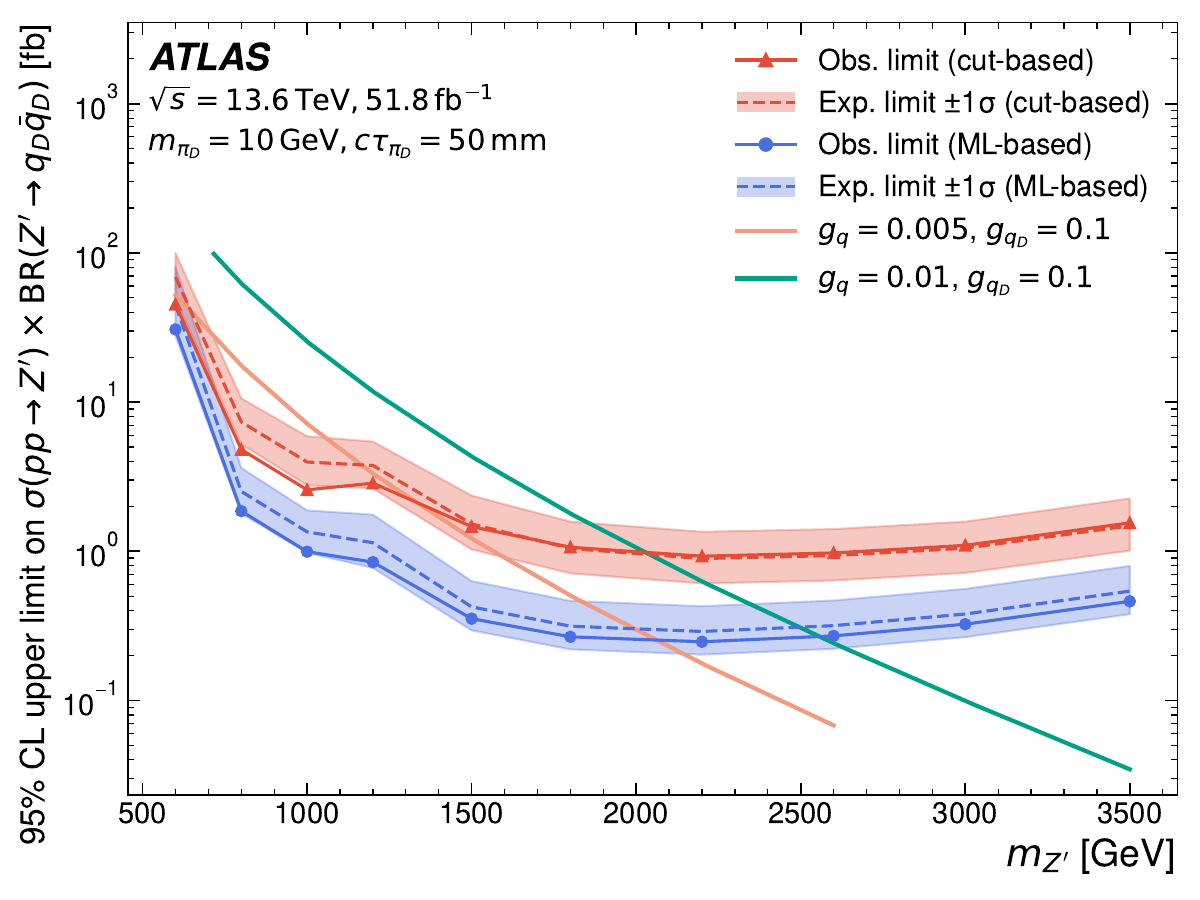}
        \caption{$s$-channel, $c\tau_{\pi_\mathrm{D}} = 50$~mm}
        \label{fig:subfigure2}
    \end{subfigure}
    \begin{subfigure}[b]{0.35\textwidth}
        \centering
        \includegraphics[width=\textwidth]{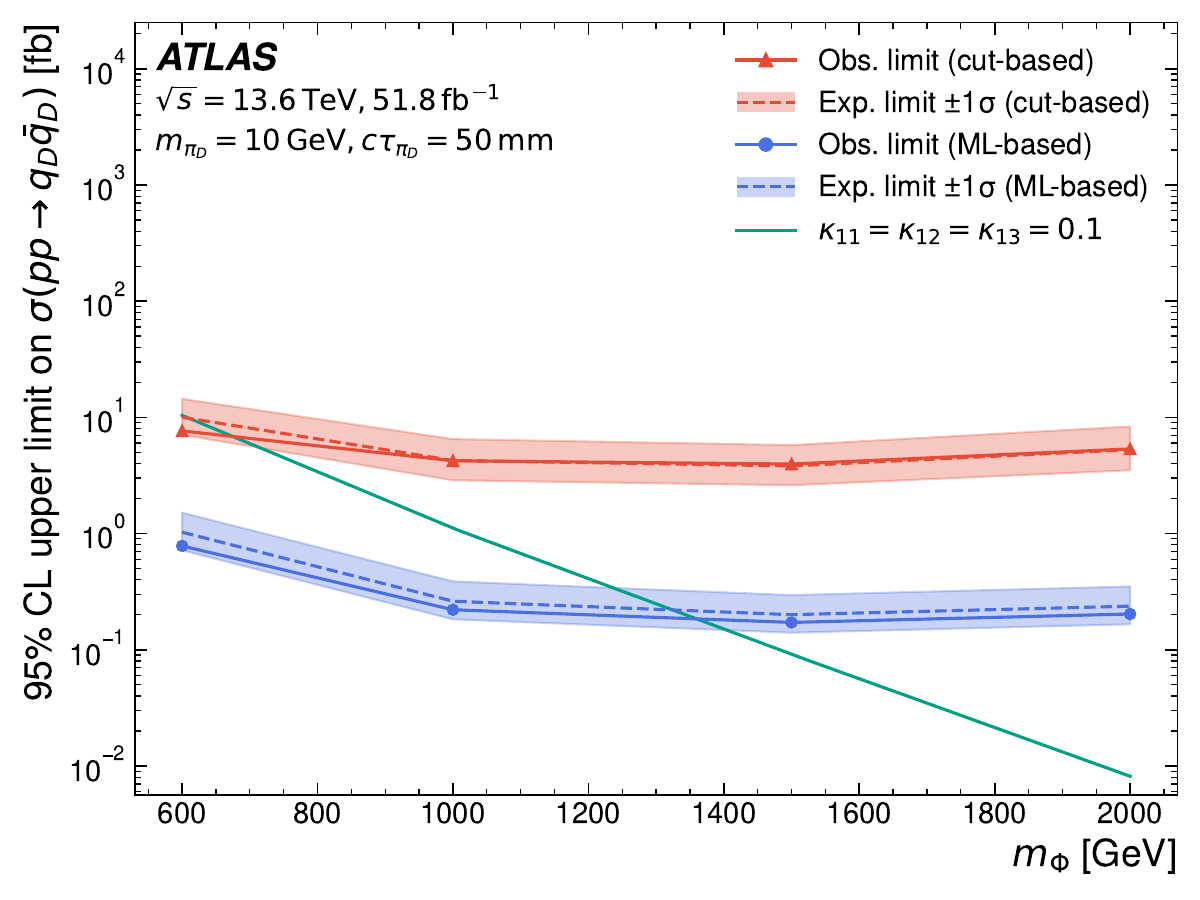}
        \caption{$t$-channel, $c\tau_{\pi_\mathrm{D}} = 50$~mm}
        \label{fig:subfigure2}
    \end{subfigure}
    \begin{subfigure}[b]{0.35\textwidth}
        \centering
        \includegraphics[width=\textwidth]{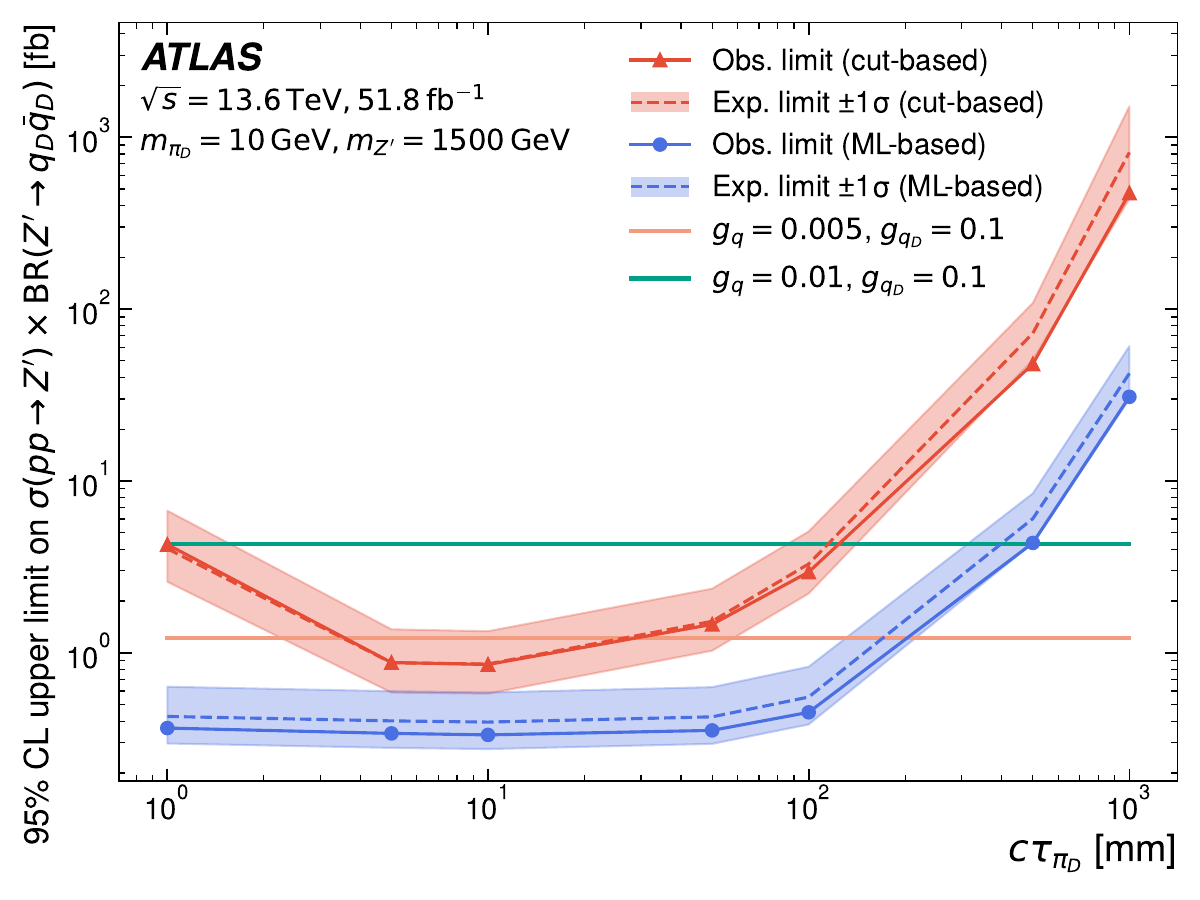}
        \caption{$s$-channel, $m_{Z'} = 1500$~GeV}
        \label{fig:subfigure2}
    \end{subfigure}
    \begin{subfigure}[b]{0.35\textwidth}
        \centering
        \includegraphics[width=\textwidth]{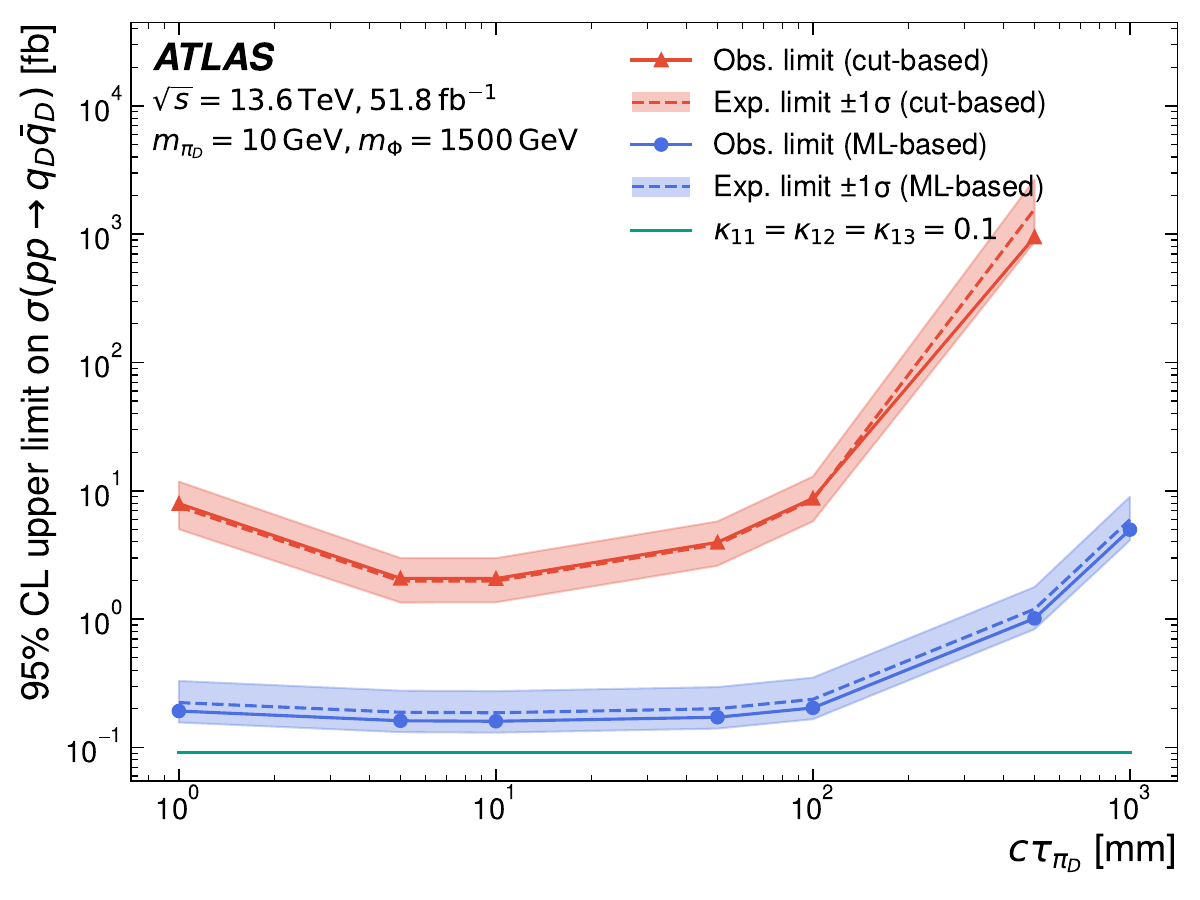}
        \caption{$t$-channel, $m_{\Phi}=1500$~GeV}
        \label{fig:subfigure2}
    \end{subfigure}
    \caption{The 95\% CL exclusion limits on (a, c) $\sigma(qq \rightarrow Z')
    \times \mathrm{Br}(Z' \rightarrow q_\mathrm{D} q_\mathrm{D})$ and (b, d) $\sigma(pp \rightarrow q_\mathrm{D} q_\mathrm{D})$
    for  $m_{\pi_\mathrm{D}}=10$~GeV \cite{EJ_ATLAS:2025bsz}. The cut-based and ML-based strategies are shown in triangles and circles, respectively. Theoretical cross-section predictions are shown as solid lines.}
    \label{fig:limits_gnn_summary}
\end{figure}

No significant excess above the
expected background is observed. Figure \ref{fig:limits_gnn_summary} shows the $95 \%$ confidence level upper limits as a function of mediator mass or dark pion lifetime $(c\tau_{\pi_\mathrm{D}})$. In the $s$-channel, limits are strongest in the high-$m_{\mathrm{jj}}$ region, while the $t$-channel, they are less dependent on the mediator mass due to a weaker correlation between the mediator mass and $m_{\mathrm{jj}}$. Overall, the ML-based approach achieves stronger exclusion limits, including in the $t$-channel scenario, which was not included in training, demonstrating the ability of the transformer based tagger to generalise to processes it was not trained on. $Z'$ masses are excluded up $2.5$ TeV  (for coupling values of $g_q = 0.01$ and $g_{q_\mathrm{D}}=0.1$), and $\Phi$ masses are excluded up to $1350$ GeV (for a quark-dark quark coupling value of $0.1$).

\section{Search for events with one displaced vertex from long-lived neutral particles decaying into hadronic jets in the muon spectrometer }

The ATLAS Collaboration searched for a single displaced vertex in the muon spectrometer (MS) \cite{1DV_MS_ATLAS:2025pak} using $140$ fb $^{-1}$ of proton–proton collision data at $\sqrt{s} = 13$ TeV, recorded during the LHC 2015–2018 data-taking period. The analysis is interpreted in the context of scalar portal, dark photon, axion-like particle (ALP), and baryogenesis models, while remaining broadly sensitive to other long-lived particle scenarios. The search is divided into two orthogonal channels, each with a dedicated trigger strategy to targeting events with and without associated prompt $Z$ boson production.

Events without an associated $Z$ boson, referred to as the ``\textit{muon-RoI}'' channel, are selected using a dedicated Muon RoI Cluster HLT trigger \cite{Muon_ROI_trigger_ATLAS:2013bsk}. This channel primarily targets scalar portal \cite{Strassler_2008} and baryogenesis \cite{baryogenesis_Cui_2015} models. In the scalar portal model, a Higgs ($H$) or alternative scalar boson ($\Phi$) produces two dark scalars ($s$), which subsequently decay to SM fermions ($\Phi / H \rightarrow ss \rightarrow f \bar{f}  f \bar{f}$). In the baryogenesis model, a weak-scale particle $\chi$ couples to the SM via a scalar $\Phi$ that mixes with the Higgs. $\chi$ is pair produced through $pp \rightarrow H \rightarrow \chi \chi$ and decays into three SM fermions, violating baryon or lepton number and generating the baryon asymmetry.

Events with an associated $Z$ boson, referred to as the ``\textit{lepton-triggered}'' channel, are selected using single- or di-lepton triggers \cite{muon_l_trig_ATLAS:2020gty, e_photon_trig_ATLAS:2019dpa} and primarily target the ALP \cite{ALP_Brivio_2017} model. In this scenario, a long-lived ALP $a$ is produced via an off-shell $Z$ boson, $pp  \rightarrow Z^* \rightarrow Za$, with $a \rightarrow  gg$, and $Z \rightarrow  l ^ + l ^ -$ ($l = e , \mu$). This channel is also sensitive to the dark photon \cite{Dark_photon_Davoudiasl:2012ag} model, where a scalar $\Phi$ decays into a $Z$ boson and a long-lived dark photon $Z_\mathrm{D}$, i.e. $ pp\rightarrow \Phi \rightarrow Z Z_\mathrm{D}$ , with $Z_\mathrm{D} \rightarrow q \bar{q}$ and $Z \rightarrow  l ^ + l ^ -$ ($l = e , \mu$). It further probes the $Z$-associated scalar portal model, which extends the original scalar portal scenario by including an additional $Z$ boson. The associated $Z$ helps reduce backgrounds, although the production cross-section is generally smaller.

The dominant background for LLP decays in the MS arises from hadronic or electromagnetic showers escaping the calorimeter, known as punch-through jets. Additional contributions come from multijet events with mismeasured energy or direction, as well as non-collision sources such as detector noise, cosmic rays, and beam-induced activity. In the muon-RoI channel, these backgrounds are suppressed through isolation requirements, thresholds on missing transverse energy and hadronic energy, and a minimum number of muon segments to reject non-collision events. In the lepton-triggered channel, the $Z$ selection significantly reduces punch-through and non-collision backgrounds, leaving $Z+$ jets as the dominant source. Consequently, looser isolation criteria are applied, along with a threshold on missing transverse energy.

Since no reliable simulation exists for all backgrounds in the MS, a data-driven ABCD method is used. The ABCD planes are defined by two multilayer perceptron (MLP) neural networks (NN) trained on distinct input features: NN1 targets DV isolation and suppression of non-collision backgrounds, while NN2 uses input features to target the DV's characteristics. Separate networks are trained for the barrel and endcap regions to account for differences in magnetic field conditions and background behaviour. Vertex reconstruction efficiency is higher in the endcaps due to straight-line tracklet fitting in the weak magnetic field, though this also reduces background rejection compared to the curvature-based reconstruction in the barrel. The NN1 and NN2 outputs define the ABCD planes used for data-driven background estimation in both detector regions.

\begin{figure}[htp]
    \centering

    \begin{subfigure}{0.32\textwidth}
        \centering
        \includegraphics[width=\linewidth]{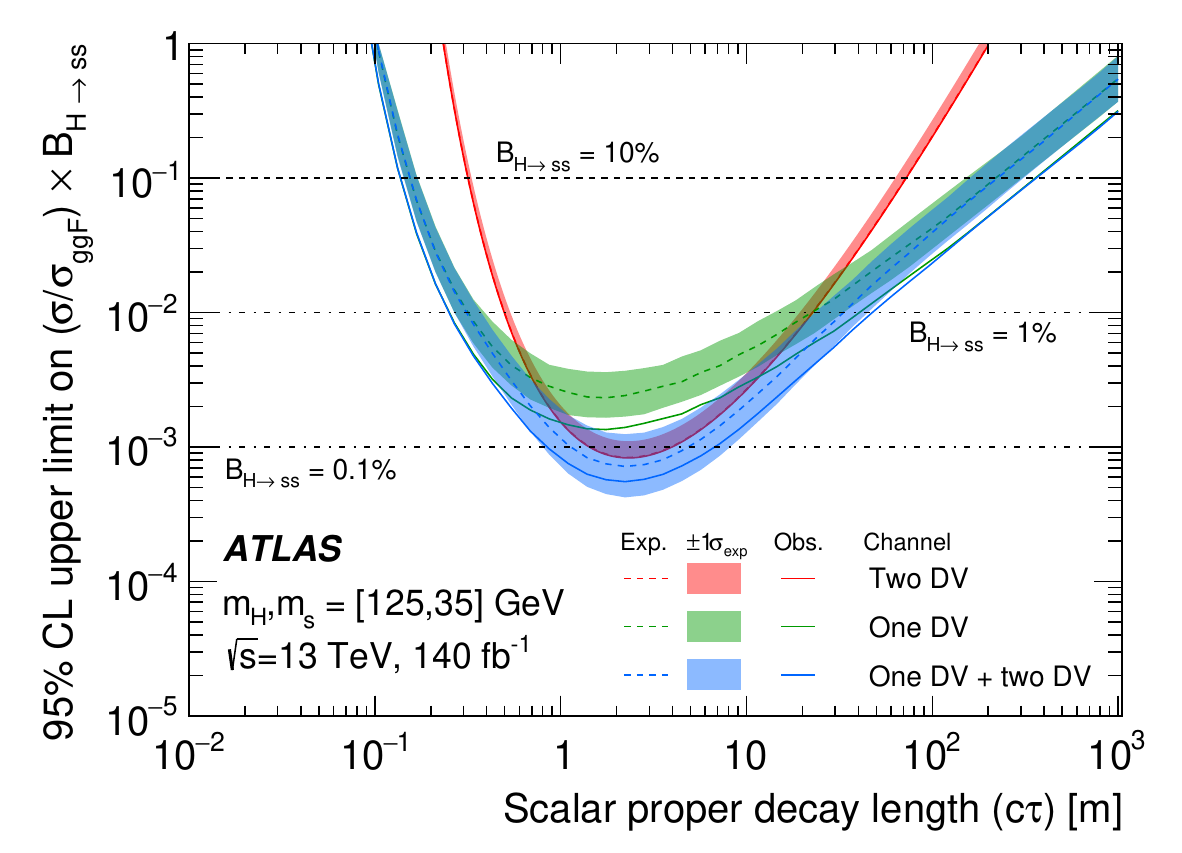}
        \caption{Scalar portal}
        \label{fig:HiggsPortal}
    \end{subfigure}
    \begin{subfigure}{0.32\textwidth}
        \centering
        \includegraphics[width=\linewidth]{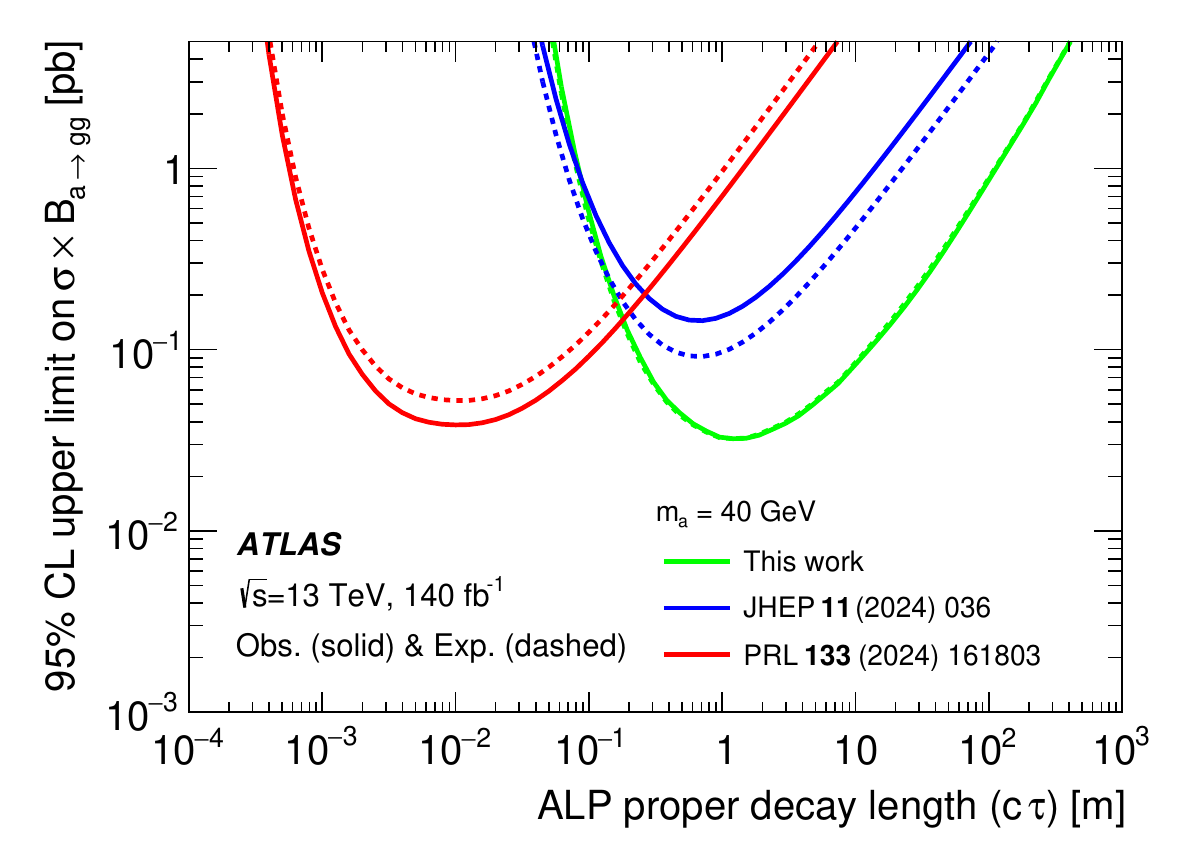}
        \caption{$Z +$ ALP}
        \label{fig:ZALP}
    \end{subfigure}
    \begin{subfigure}{0.32\textwidth}
        \centering
        \includegraphics[width=\linewidth]{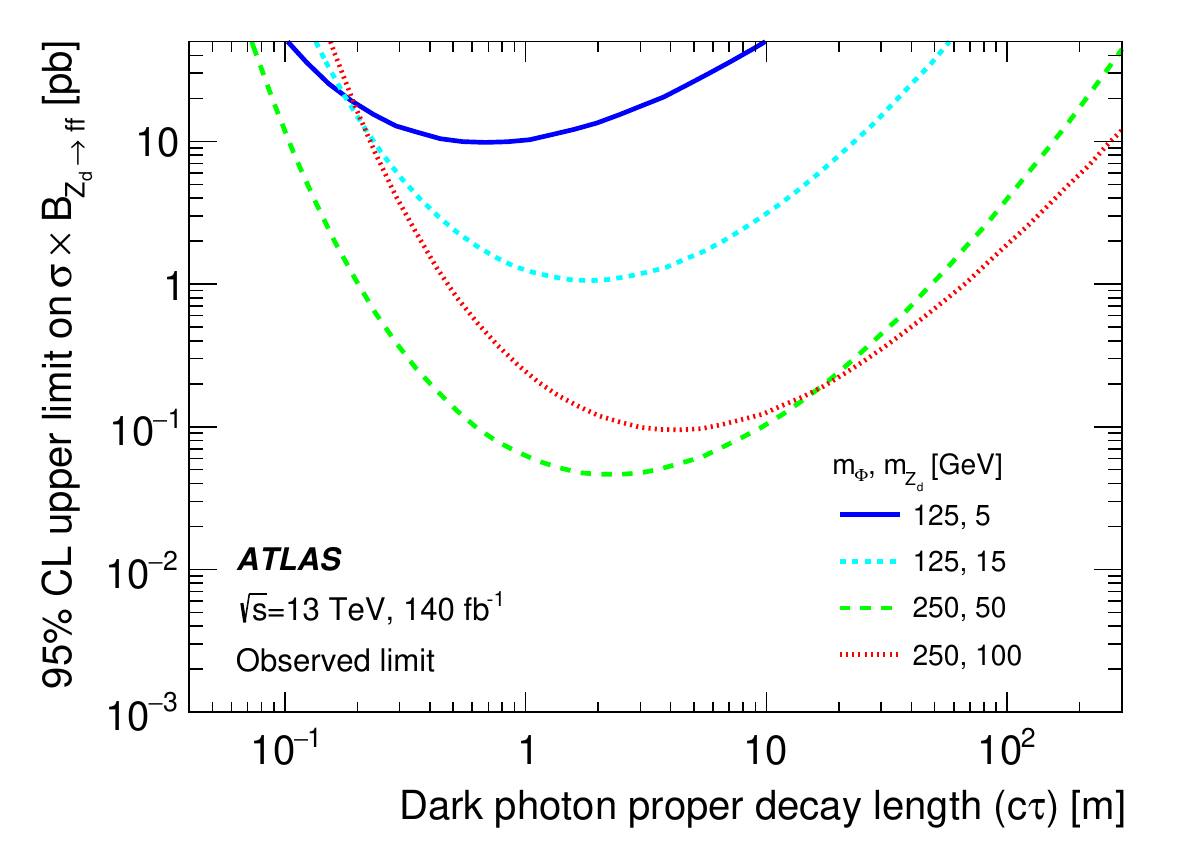}
        \caption{Dark photon}
        \label{fig:darkphoton}
    \end{subfigure} \\

    \caption{ Summary of the expected and observed $95 \%$ CL upper limits for the $\Phi / H \rightarrow ss $ model \cite{1DV_MS_ATLAS:2025pak}. (a) Comparison of the one DV \cite{1DV_MS_ATLAS:2025pak} and two DV \cite{2DV_MS_ATLAS:2022gbw} results and their combined limits on $(\sigma / \sigma_{ggH}) \times B$ for the Higgs-portal mediator with $m_s = 35$ GeV. (b) Upper limits on the $Z$+ALP production cross-section for $m_a = 40$ GeV compared with previous limits from \cite{HCal_LLP_ATLAS:2024ocv} and \cite{ID_LLP_ATLAS:2024qoo}. (c) Observed $95 \% $ CL upper limits on $\sigma \times B(Z_\mathrm{D} \rightarrow f\bar{f})$ for the dark photon $Z_\mathrm{D}$ model.}
    \label{fig:1dvtx_ms_results}
\end{figure}

No significant excess above the expected background is observed in the signal regions. Figure \ref{fig:HiggsPortal} shows the expected and observed limits from this one-DV search, the previous ATLAS two-DV search \cite{2DV_MS_ATLAS:2022gbw}, and their combination for the scalar portal benchmark with the SM Higgs boson and $m_s=35$ GeV. The two-DV search, which has very low background, provides the best sensitivity for lifetimes of $\mathcal{O}(10)$ m, while the one-DV search extends coverage to both shorter and longer lifetimes.  For the SM Higgs boson portal, branching fractions above $1 \%$ are excluded at $95 \% $ CL for LLP proper decay lengths between $5$ cm and $40$ m. Figure \ref{fig:ZALP} compares the expected and observed limits from this search with previous ATLAS displaced jet \cite{HCal_LLP_ATLAS:2024ocv} and inner-detector DV \cite{ID_LLP_ATLAS:2024qoo} searches, showing  this analysis sets the most stringent ATLAS limits to date for proper decay lengths above $\mathcal{O}(10^{-1})$ m.  Figure \ref{fig:darkphoton} shows the limits set on the cross-section times branching fraction for dark photons decaying to fermion pairs.

\section{Summary}

These proceedings summarise two recent ATLAS searches for dark sectors with unconventional signatures at the LHC. Results are presented from the first search for emerging jet pair production via a vector mediator, $Z'$, as well as for emerging jets produced through the t-channel exchange of a scalar mediator, $\Phi$. The second search targets a single displaced vertex in the muon spectrometer, improving or extending limits for the Higgs-portal and axion-like particle models compared to previous ATLAS results. The ATLAS Collaboration has ongoing efforts to continue exploring the vast phenomenological space that dark sectors present, seeking to leverage the Run-2 and forthcoming full Run-3 datasets.




\begin{thebibliography}{99}

\bibitem{Clowe_2006} D. Clowe et al., Astrophys. J. 648 (2006) L109.
\bibitem{1980Rubin_DM} V.C. Rubin et al., Astrophys. J. 238 (1980) 471.
\bibitem{Persic:1995ru} M. Persic et al., Mon. Not. Roy. Astron. Soc. 281 (1996) 27, \href{https://arxiv.org/abs/astro-ph/9506004}{arXiv:astro-ph/9506004}.
\bibitem{2018_des_collab_results} DES Collaboration, Mon. Not. Roy. Astron. Soc. 475 (2018) 3165, \href{https://arxiv.org/abs/1708.01535}{arXiv:1708.01535 [astro-ph.CO]}.
\bibitem{planck_collab_2021} Planck Collaboration, Astron.Astrophys. 652 (2021) C4, \href{https://arxiv.org/abs/1807.06209}{arXiv:1807.06209 [astro-ph.CO]}.
\bibitem{Strassler_2007} M.J. Strassler et al., Phys. Lett. B 651 (2007) 374, \href{https://arxiv.org/abs/hep-ph/0604261}{	arXiv:hep-ph/0604261}.
\bibitem{Strassler_2008} M.J. Strassler et al., Phys. Lett. B 661 (2008) 263, \href{https://arxiv.org/abs/hep-ph/0605193}{	arXiv:hep-ph/0605193}.
\bibitem{Han_2008} T. Han et al., JHEP 07 (2008) 008, \href{https://arxiv.org/abs/0712.2041}{	arXiv:0712.2041 [hep-ph]}.
\bibitem{ATLAS:2008xda} ATLAS Collaboration, JINST 3 (2008) S08003.
\bibitem{LHC_Evans:2008zzb} L. Evans et al., JINST 3 (2008) S08001, \href{}{}.
\bibitem{EJ_ATLAS:2025bsz} ATLAS Collaboration, Rept. Prog. Phys. 88 (2025) 097801, \href{https://arxiv.org/abs/2505.02429}{arXiv:2505.02429 [hep-ex]}.
\bibitem{EJ_Schwaller_2015} P. Schwaller et al., JHEP 05 (2015) 059, \href{https://arxiv.org/abs/1502.05409}{arXiv:1502.05409 [hep-ph]}.
\bibitem{ECF_Larkoski_2013} A.J. Larkoski et al., JHEP 06 (2013) 108, \href{https://arxiv.org/abs/1305.0007}{arXiv:1305.0007 [hep-ph]}.
\bibitem{atlascollaboration2025transformingjetflavourtagging} ATLAS Collaboration, \href{https://arxiv.org/abs/2505.19689}{arXiv:2505.19689 [hep-ex]} (2025).
\bibitem{1DV_MS_ATLAS:2025pak} ATLAS Collaboration, \href{https://arxiv.org/abs/2503.20445}{	arXiv:2503.20445 [hep-ex]} (2025).
\bibitem{baryogenesis_Cui_2015} Y. Cui et al., JHEP 02 (2015) 049, \href{https://arxiv.org/abs/1409.6729}{arXiv:1409.6729 [hep-ph]}.
\bibitem{ALP_Brivio_2017} I. Brivio et al., Eur. Phys. J. C 77 (2017) 5111, \href{https://arxiv.org/abs/1701.05379}{	arXiv:1701.05379 [hep-ph]}.
\bibitem{Dark_photon_Davoudiasl:2012ag} H. Davoudiasl et al., Phys. Rev. D 85 (2012) 115019, \href{https://arxiv.org/abs/1203.2947}{arXiv:1203.2947 [hep-ph]}.
\bibitem{Muon_ROI_trigger_ATLAS:2013bsk} ATLAS Collaboration, JINST 8 (2013) P07015, \href{https://arxiv.org/abs/1305.2284}{	arXiv:1305.2284 [hep-ex]}.
\bibitem{muon_l_trig_ATLAS:2020gty} ATLAS Collaboration, JINST 15 (2020) P09015, \href{https://arxiv.org/abs/2004.13447}{arXiv:2004.13447 [physics.ins-det]}.
\bibitem{e_photon_trig_ATLAS:2019dpa} ATLAS Collaboration, Eur. Phys. J. C 80 (2020) 47, \href{https://arxiv.org/abs/1909.00761}{	arXiv:1909.00761 [hep-ex]}.
\bibitem{2DV_MS_ATLAS:2022gbw} ATLAS Collaboration, Phys. Rev. D 106 (2022) 032005, \href{https://arxiv.org/abs/2203.00587}{arXiv:2203.00587 [hep-ex]}.
\bibitem{ID_LLP_ATLAS:2024qoo} ATLAS Collaboration, Phys. Rev. Lett. 133 (2024) 161803, \href{https://arxiv.org/abs/2403.15332}{	arXiv:2403.15332 [hep-ex]}.
\bibitem{HCal_LLP_ATLAS:2024ocv} ATLAS Collaboration, JHEP 11 (2024) 036, \href{https://arxiv.org/abs/2407.09183}{	arXiv:2407.09183 [hep-ex]}.

\end{thebibliography}

\end{document}